\documentstyle[12pt,epsf]{article} 

\def\lsim{\mathrel{\lower2.5pt\vbox{\lineskip=0pt\baselineskip=0pt 
           \hbox{$<$}\hbox{$\sim$}}}} 
\def\gsim{\mathrel{\lower2.5pt\vbox{\lineskip=0pt\baselineskip=0pt 
           \hbox{$>$}\hbox{$\sim$}}}}

\def\d{\delta}

\def\L{\Lambda}
\def\bL{\bar{\Lambda}}

\def\e{\epsilon}

\begin{document} 
\begin{flushright}
DPNU-02-15\\ hep-ph/0206153
\end{flushright}

\vspace{10mm}

\begin{center}
{\Large \bf 
 Cosmological constant and curved 5D geometry}

\vspace{20mm}
 Masato ITO 
 \footnote{E-mail address: mito@eken.phys.nagoya-u.ac.jp}
\end{center}

\begin{center}
{
\it 
{}Department of Physics, Nagoya University, Nagoya, 
JAPAN 464-8602 
}
\end{center}

\vspace{25mm}

\begin{abstract}
 We study the value of cosmological constant in de Sitter brane
 embedded in five dimensions with positive, vanishing and negative
 bulk cosmological constant.
 In the case of negative bulk cosmological constant, we show that
 not zero but tiny four-dimensional cosmological constant can be
 realized by tiny deviation from bulk curvature of the Randall-Sundrum
 model.
\end{abstract} 

\newpage 
%
%%%%%%%%%%%%%%%%%%%%%%%%%%%%%%%%%% 1 %%%%%%%%%%%%%%%%%%%%%%%%%%%%%%%%
 \section{Introduction}

 Cosmological constant problem is one of most important
 problems in particle physics.
 From recent sophisticated observation \cite{Perlmutter:1998np},
 our universe is acceleratingly expanding at present.   
 This implies that the universe is de Sitter, namely, 
 the four-dimensional cosmological constant $\L_{4}$ is positive. 
 According to the observed value, we have
 $\L_{4}\sim 10^{-120}M^{4}_{\rm pl} \sim (10^{-3}\;{\rm eV})^{4}
 \sim (0.1\;{\rm mm})^{-4}$.
 When we would like to explain the reason why the $\L_{4}$ is extremely 
 small, from the viewpoint of field theory, we confront a huge hierarchy
 between the cosmological constant and Planck scale.
 However it is expected that the origin of the hierarchy should be
 explained by the dynamics of underlying theory, for instance, 
 string theory or M-theory.

 As an interesting challenge to cosmological constant problem, 
 Rubakov and Shaposhnikov proposed a model that the
 vanishing four-dimensional cosmological constant can be realized in the
 framework of six-dimensional world with warped two extra dimensions
 \cite{Rubakov:1983bz}.
 Namely, the suggestion of warped extra dimensions enables Einstein
 equations to have solutions with vanishing four-dimensional
 cosmological constant.
 Based on the five-dimensional theory with an infinite extra
 dimension, an alternative proposal was done by Randall and Sundrum
 \cite{Randall:1999vf}.
 The setup is that a flat $3$-brane is embedded in
 $AdS_{5}$ background geometry.  
 In the Randall-Sundrum model, the four-dimensional cosmological
 constant in brane is equal to zero due to the fine-tuning between the
 brane tension and the bulk cosmological constant, and the localized
 gravity on the brane generates the usual Newton's law
 \cite{Ito:2001nc,Lykken:1999nb,Giddings:2000mu,Csaki:2000fc}.
 Thus the vanishing $\L_{4}$ is obtained by using extra dimension and
 fine-tuning of parameters.
 However actual value of $\L_{4}$ is not zero but very small.
 
 Taking account of accelerating universe at present, we consider
 that it is necessary to embed four-dimensional de Sitter brane
 into higher dimensional world.
 It was demonstrated that gravity can be localized on de Sitter brane
 embedded in the five-dimensional curved background
 \cite{Karch:2000ct,Miemiec:2000eq,Schwartz:2000ip,Ito:2002qp,Brevik:2002yj}.
 Thus this implies that there exists the normalizable zero mode which
 reproduces the usual four-dimensional Newton's law at large distance.
 Furthermore the extensions of Randall-Sundrum model were widely
 investigated \cite{Ito:2001fd,Ito:2002nw}.
 In the setup, we are interested in the value of the four-dimensional
 cosmological constant in de Sitter brane.

 In this paper, we consider a single de Sitter brane embedded in five
 dimensions with positive, vanishing and negative bulk cosmological
 constant.
 We show that the four-dimensional cosmological constant in brane can be
 controlled by the curvature of bulk. 
 In particular, in the case of negative bulk cosmological constant,
 we show that the observed value of four-dimensional cosmological
 constant can be realized by tiny deviation from bulk curvature of the
 Randall-Sundrum model.

 The plan of this paper is as follows.
 In section 2 we will present a setup considered here and
 we show the warp factors for positive, vanishing and negative
 bulk cosmological constant.
 In section 3 we show that the resulting volcano potential in
 Shr$\ddot{\rm o}$dinger equation for the gravitational fluctuations.
 We point out that the gravity can be localized on the
 de Sitter brane and there exists a normalizable zero mode.
 In section 4 we calculate the normalization factor of the normalizable
 zero mode and derive the four-dimensional cosmological constant
 $\L_{4}$ in brane.
 Moreover we study the behavior of $\L_{4}$ for the bulk curvature.
 Finally our conclusions are described.
%%%%%%%%%%%%%%%%%%%%%%%%%%%%%%%%% 2 %%%%%%%%%%%%%%%%%%%%%%%%%%%%%%%%%
 \section{The setup}

 We consider five-dimensional theory with a single $3$-brane and
 an infinite extra dimension $y$.
 The action with brane tension $V$ is given by
 \begin{eqnarray}
  S=\int d^{4}x\;dy\sqrt{-G}
    \left(\;\frac{M^{3}}{2}{\cal R}
    -\L\;\right)-\int d^{4}x\;\sqrt{-g^{(4)}}\;V\,,\label{eqn1}
 \end{eqnarray} 
 where $\L$ and $M$ are bulk cosmological constant and
 five-dimensional fundamental scale, respectively. 
 The ansatz for metric in five dimensions is written in the following
 form
 \begin{eqnarray}
  ds^{2}=a^{2}(y)g_{\mu\nu}dx^{\mu}dx^{\nu}+dy^{2}\equiv
 G_{MN}dx^{M}dx^{N}\,,\label{eqn2}
 \end{eqnarray}
 where $a(y)$ is warp factor and it is assumed that a fifth dimension $y$
 has  $Z_{2}$ symmetry.
 The four-dimensional slice is de Sitter as follows,
 \begin{eqnarray}
  g_{\mu\nu}dx^{\mu}dx^{\nu}=-dx^{2}_{0}+e^{2\sqrt{\bL}\;x_{0}}
  \left(dx^{2}_{1}+dx^{2}_{2}+dx^{2}_{3}\right)\,.\label{eqn3}
 \end{eqnarray}
 Here we define $\bL\equiv \L_{4}/M^{2}_{\rm pl}$, where $\L_{4}$ and 
 $M_{\rm pl}$ are four-dimensional cosmological constant and
 Planck scale, respectively.
 Solving Einstein equations of this setup, we can obtain warp factors
 for $\L>0$, $\L=0$ and $\L<0$ as follows,
 \begin{eqnarray}
  \L>0&:&a(y)=L\sqrt{\bL}\sin\frac{c-|y|}{L}\label{eqn4}\\
  \L=0&:&a(y)=\sqrt{\bL}\left(c-|y|\right)\label{eqn5}\\
  \L<0&:&a(y)=L\sqrt{\bL}\sinh\frac{c-|y|}{L}\label{eqn6}\,,
 \end{eqnarray}
 where $c$ is a positive integration constant and the curvature of bulk
 is given by
 \begin{eqnarray}
  L=\sqrt{\frac{6M^{3}}{|\L|}}\,.\label{eqn7}
 \end{eqnarray}
 Note that $c$ is the distance between the brane and the horizon.
 Imposing the normalization condition $a(0)=1$, $\bL$
 can be written in terms of $c$, namely, 
 we have $\bL=L^{-2}\sin^{-2}c/L$ for $\L>0$,
 $\bL=c^{-2}$ for $\L=0$, and $\bL=L^{-2}\sinh^{-2}c/L$ for $\L<0$.
 The delta function due to presence of a brane at $y=0$ leads to the
 relation between the brane tension $V$ and $c$.
 For $\L>0$, $\L=0$ and $\L<0$, the brane tension is expressed as
 \begin{eqnarray}
  \L>0&:&V=\frac{6M^{3}}{L}\cot\frac{c}{L}\label{eqn8}\\
  \L=0&:&V=\frac{6M^{3}}{c}\label{eqn9}\\
  \L<0&:&V=\frac{6M^{3}}{L}\coth\frac{c}{L}\label{eqn10}\,.
 \end{eqnarray}
 For each $\L$,  positive $c$ yields the brane world with positive
 tension brane.

%%%%%%%%%%%%%%%%%%%%%%%%%%%%%%%%% 3 %%%%%%%%%%%%%%%%%%%%%%%%%%%%%%%%%
 \section{Localized gravity}

 Localized gravity on the brane can be realized in setup considered
 here.
 Let us review the localization of gravity 
 \cite{Randall:1999vf,Karch:2000ct,Ito:2002qp,Brevik:2002yj}.

 By transforming from $y$ to conformally flat coordinate 
 $z=\int dy\; a^{-1}(y)$,
 the gravitational fluctuation $h_{\mu\nu}(x,z)$ around the
 background metric is governed by familiar one-dimensional
 Shr$\ddot{\rm o}$dinger equation as follows,
 \begin{eqnarray}
  \left[\;
  -\frac{d^{2}}{dz^{2}}+V(z)\;\right]\psi=m^{2}\psi\,,\label{eqn11}
 \end{eqnarray}
 where $V(z)$ is volcano potential, $\psi(z)$ corresponds to the
 fluctuation with dependence of $z$ and $m$ is four-dimensional mass.

 The conformally flat coordinate $z$ is given by  
 \cite{Karch:2000ct,Ito:2002qp}
 \begin{eqnarray}
 \L>0&:&z(y)=sgn(y)\frac{1}{\sqrt{\bL}}
 \left\{{\rm arc}\cosh\left(\frac{1}{\sin\frac{c-|y|}{L}}\right)
 -z_{0}\sqrt{\bL}\right\}\label{eqn12}\\
 \L=0&:&z(y)=sgn(y)\frac{1}{\sqrt{\bL}}
 \left\{\log\left(\frac{1}{c-|y|}\right)-
        z_{0}\sqrt{\bL}\right\}\label{eqn13}\\
 \L<0&:&z(y)=sgn(y)\frac{1}{\sqrt{\bL}}
 \left\{{\rm arc}\sinh\left(\frac{1}{\sin\frac{c-|y|}{L}}\right)
 -z_{0}\sqrt{\bL}\right\}\label{eqn14}\,,
 \end{eqnarray}
 where the constant $z_{0}$ is determined by imposing $z(y=0)=0$.
 Concretely we obtain
 \begin{eqnarray}
 \L>0&:&z_{0}=\frac{1}{\sqrt{\bL}}{\rm arc}\cosh
 \left(\frac{1}{\sin\frac{c}{L}}\right)\label{eqn15}\\
 \L=0&:&z_{0}=\frac{1}{\sqrt{\bL}}\log\frac{1}{c}\label{eqn16}\\
 \L<0&:&z_{0}=\frac{1}{\sqrt{\bL}}{\rm arc}\sinh
 \left(\frac{1}{\sin\frac{c}{L}}\right)\label{eqn17}\,.
 \end{eqnarray}
 Thus the conformally flat coordinate $z$ runs from $-\infty$ to $+\infty$.

 For $\L>0$, $\L=0$ and $\L<0$, the resulting volcano potential in the
 Shr$\ddot{\rm o}$dinger equation is explicitly expressed as
 \cite{Karch:2000ct,Ito:2002qp}
 \begin{eqnarray}
 \L>0&:& V(z)=\frac{9}{4}\bL-\frac{15}{4}
 \frac{\bL}{\cosh^{2}\sqrt{\bL}\left(z_{0}+|z|\right)}
 -3\sqrt{\bL}\tanh z_{0}\sqrt{\bL}\;\d(z)\label{eqn18}\\
 \L=0&:& V(z)=\frac{9}{4}\bL-3\sqrt{\bL}\d(z)\label{eqn19}\\
 \L<0&:& V(z)=\frac{9}{4}\bL+\frac{15}{4}
 \frac{\bL}{\sinh^{2}\sqrt{\bL}\left(z_{0}+|z|\right)}
 -3\sqrt{\bL}\coth z_{0}\sqrt{\bL}\;\d(z)\label{eqn20}\,.
 \end{eqnarray}
 Note that the wave function depends on both the form of the volcano
 potential and the range of eigenvalue $m^{2}$.
 Furthermore there no exist tachyon's modes $m^{2}<0$ since
 the Hamiltonian in Eq.(\ref{eqn11}) is positive definite value
 \cite{Csaki:2000fc}.
 As described in Eqs.(\ref{eqn18})$-$(\ref{eqn20}),
 The localization of gravity occurs due to an attractive
 force potential of delta function at origin. 

 For $\L=0$, the potential is positive constant $9\bL/4$
 everywhere except at origin.
 For $\L>0$ and $\L<0$, the potential goes to a constant $9\bL/4$
 for $|z|\rightarrow \infty$.
 Solving Shr$\ddot{\rm o}$dinger equation for $0\leq m^{2}< 9\bL/4$,
 the wave function of $\L=0$ is expressed in terms of exponential function
 while the wave functions of $\L>0$ and $\L<0$ can be written in terms of
 hypergeometric function \cite{Ito:2002qp,Brevik:2002yj}.
 Taking account of the delta function at origin and the finiteness
 of wave function, 
 there exists only the wave function of $m^{2}=0$ for each $\L$.
 Up to the normalization factor, wave functions of zero mode $(m^{2}=0)$
 can be taken the following forms \cite{Ito:2002qp}
 \begin{eqnarray}
  \L>0&:& \psi_{0}\sim \cosh^{-3/2}\sqrt{\bL}(z_{0}+|z|)\,.\label{eqn21}\\
  \L=0&:& \psi_{0}\sim e^{-\frac{3}{2}\sqrt{\bL}|z|}\label{eqn22}\\
  \L<0&:& \psi_{0}\sim \sinh^{-3/2}\sqrt{\bL}(z_{0}+|z|)\label{eqn23}\,.
 \end{eqnarray}
 Thus there exists a normalizable zero mode for each $\L$.
 On the other hand, the wave functions of $m^{2}\geq 9\bL/4$ become
 plane waves with continuous modes.
 Consequently, there is a mass gap between the zero mode and massive
 continuous modes, and the beginning of massive mode starts from $9\bL/4$.
 The normalizable zero mode wave function can produce the
 four-dimensional Newton's law. 
 At large distance the small correction to Newton's law is generated
 by the contributions of wave functions with continuous modes.

%%%%%%%%%%%%%%%%%%%%%%%%%%%% 4 %%%%%%%%%%%%%%%%%%%%%%%%%%%%%%%%%%%%%%%%%
 \section{Cosmological constant in brane}

 Using the result obtained in previous section, we can obtain
 the four-dimensional cosmological constant $\L_{4}$.
 Since the gravitational potential between two unit masses
 at distance $r$ via zero mode is expressed as
 $V(r)=M^{-3}|\psi_{0}(0)|^{2}/r$, the Planck scale is
 given by $M^{-2}_{\rm pl}=M^{-3}|\psi_{0}(0)|^{2}$.

 Performing the normalization integral of zero mode wave function
 in Eqs.(\ref{eqn21})$-$(\ref{eqn23}),
 then we can obtain the normalization factor for $\L>0$, $\L=0$ and
 $\L<0$.
 Using the normalization factor including  $\bL$, we can get the
 four-dimensional cosmological constant in de Sitter brane as follows
 \begin{eqnarray}
 \L>0&:&\L_{4}= f(x)\;M^{4}_{\rm pl}\left(\frac{M}{M_{\rm pl}}\right)^{6}
 \label{eqn24}\\
 \L=0&:&\L_{4}= \frac{4}{9}\;
 M^{4}_{\rm pl}\left(\frac{M}{M_{\rm pl}}\right)^{6}\label{eqn25}\\
 \L<0&:&\L_{4}= g(x)\;M^{4}_{\rm pl}\left(\frac{M}{M_{\rm pl}}\right)^{6}
 \label{eqn26} 
 \end{eqnarray}
 where
 \begin{eqnarray}
 f(x)&=&
 (x^{2}+1)^{3}\left(\frac{\pi}{2}-\frac{x}{x^{2}+1}
 -{\rm arc}\tan x\right)^{2}\label{eqn27}\\
 g(x)&=&(x^{2}-1)^{3}\left(\frac{x}{x^{2}-1}
 -\frac{1}{2}\log\frac{x+1}{x-1}\right)^{2}\label{eqn28}
 \end{eqnarray}
 Here we used $\bL= \L_{4}/M^{2}_{\rm pl}$.
 Moreover we introduced a dimensionless parameter as follows
 \begin{eqnarray}
 x\equiv \frac{LV}{6M^{3}}\label{eqn29}\,.
 \end{eqnarray} 
%
%%%%%%%%%%%%%%%%%%%% fig.1 %%%%%%%%%%%%%%%%%%%%%%%%%%%%
\begin{figure}
      \epsfxsize=8cm
\centerline{\epsfbox{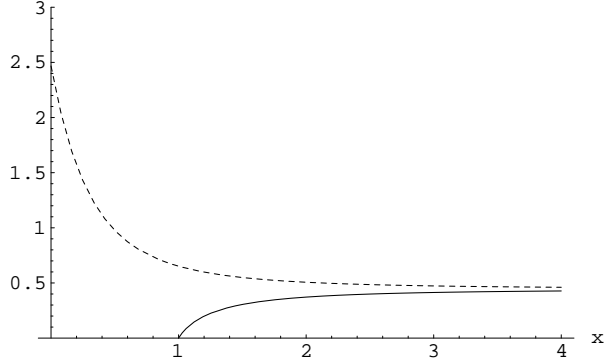}}
\caption{overall factor for $\L>0$ (dash) and $\L<0$ (solid).}
\label{fig1} 
\end{figure}
%%%%%%%%%%%%%%%%%%%%%%%%%%%%%%%%%%%%%%%%%%%%%%%%%%%%%%%
 Note that $x$ is determined by the bulk curvature $L$ which
 is fixed by bulk cosmological constant $\L$.
 As described in Eqs.(\ref{eqn24})$-$(\ref{eqn26}),
 we are interested in overall factor of
 $M^{4}_{\rm pl}(M/M_{\rm pl})^{6}$ for each $\L$.
 This is because it is possibility that the factor may generate not zero
 but tiny value of $\L_{4}$.
 Let us investigate the value of overall factor for $\L=0$, $\L>0$
 and $\L<0$, separately.

 For $\L=0$, from Eq.(\ref{eqn25}), the observed value of $\L_{4}$
 leads to the ratio $M/M_{\rm pl}\sim 10^{-20}$.
 Namely, we have $M={\cal O}(100\;{\rm MeV})$.
 Accidentally, the five-dimensional fundamental scale is closely
 equal to the QCD scale.
 This implies that in the case of de Sitter brane embedded in $\L=0$
 background, the observed value of the four-dimensional
 cosmological constant cannot be phenomenologically obtained.

 In the case of $\L>0$ and $\L<0$,
 $\L_{4}$ is obviously determined by $x$ defined in Eq.(\ref{eqn29}).
 It is necessary to study the behavior of $\L_{4}$ for the variation
 of $x$.

 For $\L>0$, the overall factor $f(x)$, which determines the value of
 $\L_{4}$, is shown in Figure.\ref{fig1}(dash curve).
 Since the limit of $x\rightarrow \infty$ yields $f(x)\rightarrow 4/9$,
 we have $4/9\leq f(x)\leq \pi^{2}/4$ for $x\geq 1$.
 Looking the relation between $x$ and background geometry,
 the transition from small $x$ to large $x$ corresponds to the
 continuous transition from $\L>0$ background to $\L=0$ background.
 As for warp factor, from Eqs.(\ref{eqn4}) and (\ref{eqn5}),
 the warp factor of $\L>0$ goes to the that of $\L=0$.
 For arbitrary $x$, the magnitude of $f(x)$ is almost order of the
 unity.
 In similar to the case of $\L=0$, 
 we must take $M={\cal O}(100\;{\rm MeV})$ in order to obtain the
 observed value of the four-dimensional cosmological constant.
 Thus in the case of $\L>0$ background we cannot obtain extremely tiny
 $\L_{4}$. 

 For $\L<0$, the overall factor $g(x)$ is shown as solid curve
 in Figure.\ref{fig1}.
 From Eq.(\ref{eqn10}), we have $x\geq 1$.
 Taking the limit of $x\rightarrow\infty$, we have $g(x)\rightarrow 4/9$
 and Eq.(\ref{eqn26}) goes to Eq.(\ref{eqn25}).
 Furthermore, from Eqs.(\ref{eqn5}) and (\ref{eqn6}), the warp factor
 of $\L<0$ goes to the that of $\L=0$. 
 This implies the transition from $\L<0$ to $\L=0$.
 Particularly the case of $x=1$ corresponds to a flat
 $3$-brane $(\L_{4}=0)$, obviously, this situation is as same as the
 Randall-Sundrum model of Ref.\cite{Randall:1999vf}. 
 The point at $x=1$ leads to the bulk curvature of the Randall-Sundrum
 model as follows,
 \begin{eqnarray}
  L_{\rm RS}=\frac{6M^{3}}{V}\,.\label{eqn30}
 \end{eqnarray}
 From Eq.(\ref{eqn6}) and $a(0)=1$, in the limit of $x\rightarrow 1$
 we obtain $a(y)\rightarrow e^{-|y|/L_{\rm RS}}$ which is consistent with
 warp factor of the Randall-Sundrum model.
 In order to get not zero but tiny $\L_{4}$, we consider that
 the curvature of bulk has tiny deviation from $L_{\rm RS}$.
 Namely, we take $L=L_{\rm RS}+l\;(l\ll 1)$,
 from Eq.(\ref{eqn29}),
 $x=1+\epsilon\;(\epsilon\equiv l/L_{\rm RS}\ll 1)$ is obtained. 
 Using $g(1+\epsilon)\sim \epsilon$, the four-dimensional
 cosmological constant in de Sitter brane is given by
 \begin{eqnarray}
 \L_{4}\sim\epsilon M^{4}_{\rm pl}
 \left(\frac{M}{M_{\rm pl}}\right)^{6}\,.\label{eqn31}
 \end{eqnarray}
 As expected in theories with extra dimensions and in future collider LHC,
 adopting $M= {\rm TeV}$ and observed value of $\L_{4}$,  
 we have $\e\sim 10^{-24}$.
 If $L_{\rm RS}= 0.1\;{\rm mm}$ is close to the lower bound at which
 gravitational experiments are performed in the future \cite{Hoyle:2000cv},
 then $l\sim 10^{-13}\;{\rm fm}$.
 Thus we point out that, in de Sitter brane embedded in five dimensions
 with negative bulk cosmological constant,
 it is possibility that not zero but extremely tiny $\L_{4}$ comes from
 the extremely tiny deviation from bulk curvature of Randall-Sundrum model. 

%%%%%%%%%%%%%%%%%%%%%%%%%%%%%%%%%%%%%%%%%%%%%%%%%%%%%%%%%%%%%%%%%%%%%
 \section{Conclusion}
 We are interested in the recent observation that the four-dimensional
 cosmological constant $\L_{4}$ is not zero but positive tiny constant.
 We considered the de Sitter brane embedded in five dimensions with
 positive, vanishing and negative bulk cosmological constant $\L$.
 In the setup for each $\L$, gravity can be localized on the brane and
 there exists a normalizable zero mode which generates the usual
 Newton's law at large distance.
 By calculating the normalization integral of a zero mode,
 we derived $\L_{4}$ for each $\L$.
 In the case of $\L>0$ and $\L=0$, fundamental scale in five dimensions
 must be taken MeV scale in order to obtain the observed value of
 $\L_{4}$.
 For $\L<0$, we pointed out that the observed value of $\L_{4}$ can be
 realized by tiny deviation from bulk curvature of the Randall-Sundrum
 model.
 From the viewpoint of the cosmological constant, we should consider
 the system of a de Sitter brane in $\L<0$ background geometry.
 Here we mention some comments.
 In this paper it is not saying that we solved the cosmological constant
 problem.
 Since we considered the bulk cosmological constant as a parameter,
 we must explain the reason why the bulk curvature is vicinity of bulk
 curvature of the Randall-Sundrum model.
 Namely the bulk curvature is determined by the bulk cosmological
 constant governed by the fundamental theory of high energy region,
 it is unknown whether a scenario considered here can be realized by
 the dynamics of fundamental theory. 
 However we suggested a proposal to obtain not zero but extremely tiny
 four-dimensional cosmological constant without fine-tuning of couplings.
 In cosmological context a brane model presented here will be discussed
 in future.

%%%%%%%%%%%%%%%%%%%%%%%%%%%% reference %%%%%%%%%%%%%%%%%%%%%%%%%
%


\begin{thebibliography}{99}
%
 %\cite{Perlmutter:1998np}
 \bibitem{Perlmutter:1998np}
 S.~Perlmutter {\it et al.}  [Supernova Cosmology Project Collaboration],
 ``Measurements of Omega and Lambda from 42 High-Redshift Supernovae,''
 Astrophys.\ J.\  {\bf 517}, 565 (1999) [astro-ph/9812133].
%
  %\cite{Rubakov:1983bz}
 \bibitem{Rubakov:1983bz}
 V.~A.~Rubakov and M.~E.~Shaposhnikov,
 ``Extra Space-Time Dimensions: Towards A Solution To The Cosmological
 Constant Problem,''
 Phys.\ Lett.\ B {\bf 125}, 139 (1983).
%
 %\cite{Randall:1999vf}
 \bibitem{Randall:1999vf}
 L.~Randall and R.~Sundrum,
 ``An alternative to compactification,''
 Phys.\ Rev.\ Lett.\  {\bf 83}, 4690 (1999) [hep-th/9906064].
%
 %\cite{Ito:2001nc}
 \bibitem{Ito:2001nc}
 M.~Ito,
 ``Newton's law in braneworlds with an infinite extra dimension,''
 Phys.\ Lett.\ B {\bf 528}, 269 (2002) [hep-th/0112224].
%
 %\cite{Lykken:1999nb}
 \bibitem{Lykken:1999nb}
 J.~Lykken and L.~Randall,
 ``The shape of gravity,'' JHEP {\bf 0006}, 014 (2000) [hep-th/9908076].
%
 %\cite{Giddings:2000mu}
 \bibitem{Giddings:2000mu}
 S.~B.~Giddings, E.~Katz and L.~Randall,
 ``Linearized gravity in brane backgrounds,'' JHEP {\bf 0003}, 023 (2000)
 [hep-th/0002091].
%
 %\cite{Csaki:2000fc}
 \bibitem{Csaki:2000fc}
 C.~Csaki, J.~Erlich, T.~J.~Hollowood and Y.~Shirman,
 ``Universal aspects of gravity localized on thick branes,''
 Nucl.\ Phys.\ B {\bf 581}, 309 (2000) [hep-th/0001033].
%
 %\cite{Karch:2000ct}
 \bibitem{Karch:2000ct}
 A.~Karch and L.~Randall,
 ``Locally localized gravity,'' JHEP {\bf 0105}, 008 (2001)
 [hep-th/0011156].
%
  %\cite{Miemiec:2000eq}
 \bibitem{Miemiec:2000eq}
 A.~Miemiec,
 ``A power law for the lowest eigenvalue in localized massive gravity,''
 Fortsch.\ Phys.\  {\bf 49}, 747 (2001) [hep-th/0011160].
%
  %\cite{Schwartz:2000ip}
 \bibitem{Schwartz:2000ip}
 M.~D.~Schwartz,
 ``The emergence of localized gravity,'' Phys.\ Lett.\ B {\bf 502}, 223 (2001)
 [hep-th/0011177].
%
 %\cite{Ito:2002qp}
 \bibitem{Ito:2002qp}
 M.~Ito,
 ``Localized gravity on de Sitter brane in five dimensions,''
 [hep-th/0204113].
%
 %\cite{Brevik:2002yj}
 \bibitem{Brevik:2002yj}
 I.~Brevik, K.~Ghoroku, S.~D.~Odintsov and M.~Yahiro,
 ``Localization of gravity on brane embedded in AdS(5) and dS(5),''
 [hep-th/0204066].
%
 %\cite{Ito:2001fd}
 \bibitem{Ito:2001fd}
 M.~Ito,
 ``Warped geometry in higher dimensions with an orbifold extra dimension,''
 Phys.\ Rev.\ D {\bf 64}, 124021 (2001) [hep-th/0105186].
%
%
 %\cite{Ito:2002nw}
 \bibitem{Ito:2002nw}
 M.~Ito,
 ``Linearized gravity in flat braneworlds with anisotropic brane tension,''
 [hep-th/0202166].
% 
 %\cite{Hoyle:2000cv}
 \bibitem{Hoyle:2000cv}
 C.~D.~Hoyle, U.~Schmidt, B.~R.~Heckel, E.~G.~Adelberger,
 J.~H.~Gundlach, D.~J.~Kapner and H.~E.~Swanson,
 ``Sub-millimeter tests of the gravitational inverse-square law: 
 A search  for 'large' extra dimensions,''
 Phys.\ Rev.\ Lett.\  {\bf 86}, 1418 (2001) [hep-ph/0011014].
\end{thebibliography}
\end{document}